\documentclass[conference,letterpaper]{IEEEtran}

\IEEEoverridecommandlockouts

\usepackage[
  top=0.710in,
  bottom=1.080in,
  left=0.673in,
  right=0.673in
]{geometry}

\usepackage{cite}             

\usepackage{amsmath}  
\usepackage{amsfonts,amssymb}
\usepackage{mleftright}       
\mleftright                   

\usepackage{graphicx}         
\usepackage{booktabs}         

\usepackage[lined,boxed,commentsnumbered,linesnumbered,ruled,vlined]{algorithm2e}
\SetKwIF{MyIf}{MyElseIf}{MyElse}{if}{then}{else if}{else}{}

\usepackage[caption=false,font=footnotesize]{subfig}

\usepackage{pgfplots}
\usepackage{tikz}
\usetikzlibrary{shapes,patterns,decorations.text,decorations.pathreplacing,calc}
\pgfplotsset{compat=1.8}
\usetikzlibrary{external}
\usepackage{pdfpages}

\usepackage{nicefrac}  
\usepackage{comment}
\usepackage{balance}

\usepackage[capitalize]{cleveref} 
	\crefname{equation}{}{}
	\crefname{theorem}{Theorem}{Theorems}
	\crefname{lemma}{Lemma}{Lemmas}
	\crefname{cor}{Corollary}{Corollaries}
	\crefname{prop}{Proposition}{Propositions}
	\crefname{note}{Note}{Notes}
	\crefname{appsec}{Appendix}{Appendices}
	\crefname{definition}{Definition}{Definitions}
	\crefname{conj}{Conjecture}{Conjectures}
	\crefname{construction}{Construction}{Constructions}
	
\usepackage[dvipsnames]{xcolor}
\renewcommand{\vec}[1]{\boldsymbol{#1}} 

\newcommand{\Reals}{\mathbb{R}}      

\newcommand{\Prv}[1]{\Pr\left[#1\right]}

\newcommand{\EE}[2][]{%
	\ifthenelse{\equal{#1}{}}%
	{\mathbb{E} \left[ #2 \right]}
	{\mathbb{E}_{#1} \left[ #2 \right]}
}

\newcommand{\nEE}[2][]{%
	\ifthenelse{\equal{#1}{}}%
	{\mathbb{E} \left\lVert #2 \right\rVert}
	{\mathbb{E}_{#1} \left\lVert #2 \right\rVert}
}
\newcommand{\nsEE}[2][]{%
	\ifthenelse{\equal{#1}{}}%
	{\mathbb{E} \left\lVert #2 \right\rVert^2}
	{\mathbb{E}_{#1} \left\lVert #2 \right\rVert^2}
}

\newcommand{\NN}[2]{\mathcal{N}\left({#1},{#2}\right)} 

\newcommand{\BC}[3]{\mathcal{BC}\left(#1,#2,#3\right)}

\newcommand{\set}[1]{\mathcal{#1}}

\newcommand{\numa}{M}

\newcommand{\todo}[1]{\textbf{TODO: #1}}

\newcommand{\sigmaDP}{\sigma_{\mathrm{DP}}}
\newcommand{\betaDP}{\beta_{\mathrm{DP}}}

\newcommand{\tildebar}[1]{\tilde{\bar{#1}}}
\newcommand{\tildehat}[1]{\tilde{\hat{#1}}}

\newtheorem{theorem}{Theorem}
\newtheorem{cor}{Corollary}
\newtheorem{example}{Example}
\newtheorem{lemma}{Lemma}
\newtheorem{definition}{Definition}
\newtheorem{proposition}{Proposition}

\usepackage{bm}
\usepackage{dsfont}     

\definecolor{deepmagenta}{rgb}{0.80,0.0,0.80}
\definecolor{darkmagenta}{rgb}{0.55,0.0,0.55}
\definecolor{GreenB}{rgb}{0.27,0.67,0.42}
\definecolor{darkgreen}{rgb}{0, 0.5, 0}
\definecolor{darkyellow}{rgb}{0.61,0.53,0.05}
\definecolor{quinacridonemagenta}{rgb}{0.6016,0.0664,0.3086}
\definecolor{lightblue}{rgb}{0.61,0.87,1.0}
\definecolor{darkred}{rgb}{0.55, 0.0, 0.0}
\definecolor{deepcarmine}{rgb}{0.66, 0.13, 0.24}
\definecolor{flamingopink}{rgb}{0.99, 0.56, 0.67}

\begin{document}

\title{Communication-Constrained Private Decentralized Online Personalized Mean Estimation} 

\author{\IEEEauthorblockN{Yauhen Yakimenka\IEEEauthorrefmark{1}, Hsuan-Yin Lin\IEEEauthorrefmark{2}, Eirik Rosnes\IEEEauthorrefmark{2}, and J{\"o}rg~Kliewer\IEEEauthorrefmark{1}}
\IEEEauthorblockA{\IEEEauthorrefmark{1}Helen and John C.~Hartmann Department of Electrical and Computer Engineering,\\ New Jersey Institute of Technology, Newark, New Jersey 07102, USA
}\IEEEauthorblockA{\IEEEauthorrefmark{2}Simula UiB, N-5006 Bergen, Norway
}
\thanks{This work was in part supported by US NSF grants 2107370 and 2201824, and the Research Council of Norway (RCN) under the PeerL project (grant 355124).}
}

\maketitle

\begin{abstract}
  We consider the problem of communication-constrained collaborative personalized mean estimation under a privacy constraint in an environment of several agents continuously receiving data according to arbitrary unknown agent-specific distributions. A consensus-based algorithm is studied under the framework of differential privacy in order to protect each agent's data.  We give a theoretical convergence analysis of the proposed consensus-based algorithm for any bounded unknown distributions on the agents’ data, showing that collaboration provides faster convergence than a fully local approach where agents do not share data, under an oracle decision rule and under some restrictions on the  privacy level and the agents' connectivity, which  illustrates the benefit of private collaboration in an online setting under a communication restriction on the agents. The theoretical faster-than-local convergence guarantee is backed up by several numerical results.
\end{abstract}

\section{Introduction}

The interest in collaborative learning has grown considerably recently, fueled by prominent frameworks such as federated learning (FL)  \cite{McMahanMooreRamageHampsonArcas17_1,Konecy-etal16_1,LiSahuTalwalkarSmith20_1}, which offers a partially decentralized approach, and fully decentralized methods like swarm learning \cite{Warnat-Herresthal2021}. A key challenge in such environments is that individual learning agents may possess distinct goals, with heterogeneous and task-specific datasets. Nevertheless, collaboration can substantially speed up learning when agents share even a limited set of common objectives. Therefore, a critical component of any collaborative algorithm for personalized learning is the ability to identify agents whose data originates from similar distributions, especially in dynamic online settings where data arrives continuously.

Personalized approaches to FL have emerged in order to develop personalized models, designed to better align with the data distributions of individual agents, see, e.g.,  \cite{Smith2017,Vanhaesebrouck2017,Hanzely2020}. Many personalized FL methods group agents into clusters to train tailored models, see, e.g.,  \cite{Ghosh2020,Fallah2020,Li2021,Marfoq2021,Ding2022,Muller2021,Even2022}. The ideal is to cluster agents with similar optimal local models, but since these models are unknown, model learning and cluster identification are intertwined. Several similarity measures have been proposed in this respect (see, e.g., \cite{Ghosh2020,Muller2021}), while other works (see, e.g., \cite{Ding2022,Even2022}) assume some a priori information on the intra-distance among the data distributions. Estimating these intra-distances are known in the literature to be a difficult task and remains a largely unsolved problem \cite[Sec.~6]{Even2022}.

This paper delves into the related problem of collaborative online personalized mean estimation, initially formulated in \cite{AsadiBelletMaillardTommasi22_1}. Here, each agent continuously receives data from its own unknown distribution. We operate in a fully decentralized setting, without a central server, distinguishing our work from FL. Furthermore, unlike FL's typical reliance on stochastic gradient descent, we concentrate on the statistical problem of mean estimation. The aim for each agent is to quickly achieve an accurate estimate of its underlying distribution mean. Following \cite{AsadiBelletMaillardTommasi22_1}, we assume an unknown class structure where agents belonging to the same class share the same data distribution mean. This problem was also considered in \cite{galante2024scalable} under a communication restriction,  i.e., there is an underlying communication graph that restricts the communication between agents, and with a privacy constraint in \cite{Yakimenka2023,yauhen2024arxiv}, but with no restrictions on the agent-to-agent communication. For the case with an underlying communication restriction, the effect of a data privacy constraint has so far not been studied, and  in this work, we address this gap by extending our work in  \cite{Yakimenka2023,yauhen2024arxiv}. The proposed solution is based  on the consensus algorithm introduced in \cite{galante2024scalable}, coupled with the concept of differential privacy (DP) \cite{Dwork06_1,DworkMcSherryNissimSmith06_1} and a new decision rule. Our main result is a theoretical convergence analysis showing that collaboration indeed provides faster convergence than a fully local approach where agents do not share data, under an oracle decision rule and under some restrictions on the privacy level and the agents’ connectivity,  for any bounded unknown distributions on the agents’ data (see Corollary~\ref{cor:sigmaDP-bound}). The theoretical faster-than-local convergence guarantee is backed up by several numerical results.

\section{Preliminaries}

\subsection{Notation}
\label{sec:notation}

In general, but with some exceptions, we use uppercase and lowercase letters for random variables (RVs) and their realization, %
respectively, and italics for sets, %
e.g., \(X\), \(x\), and \(\mathcal{X}\) 
represent a RV, its realization, and a set, respectively. %
The expectation of a RV $X$ is denoted by $\EE{X}$. %
We define $[n] \triangleq \{1,2,\dotsc, n\}$, %
while $\mathbb{N}$ denotes  the natural numbers and  
$\Reals$  the real numbers. %
$\mathcal{L}\left(\mu,b \right)$ denotes the Laplace distribution with mean $\mu$ and scale parameter $b$ (variance is $2b^2$). $X \sim \set P$ denotes that $X$ is distributed according to the distribution $\set P$. %
Standard order notations $O(\cdot)$ and $o(\cdot)$ are used for asymptotic results. %

\subsection{Differential Privacy}
We start by defining the concept of DP.

\begin{definition}
  \label{def:DP}
  A randomized function $F\colon\set{X}^n\to\set{Y}$ is $\epsilon$-differentially private if for all subsets $\set{S}\subseteq\set{Y}$ and for all $(x_1,\dotsc,x_n)\in\set{X}^n$ and $(x'_1,\dotsc,x'_n)\in\set{X}^n$ which differ in a single component, i.e., $x_i \neq x_i'$ for exactly one $i \in [n]$,
  \begin{IEEEeqnarray*}{c}
    \Prv{F(x_1,\dotsc,x_n) \in \set{S}} \leq \mathrm{e}^\epsilon \Prv{F(x'_1,\dotsc,x'_n) \in \set{S}}.
  \end{IEEEeqnarray*}
\end{definition}

\begin{lemma}
  \label{lem:DP_Laplace}
  Let $(x_1,\dotsc,x_n) \in \mathcal{X}^n$ where $\mathcal{X} = [\mu-L,\mu+L]$ for some finite values $\mu$ and $L$.  %
  Then, the noise-corrupted sample mean 
    $\nicefrac{(x_1+\cdots+x_n)}{n} + \nicefrac Zn$,
  where $Z \sim \mathcal{L}\left(0,\nicefrac{\sigmaDP}{\sqrt{2}} \right)$ and $\sigmaDP^2 \triangleq \nicefrac{8L^2}{\epsilon^2}$ is 
  $\epsilon$-differentially private for $\epsilon > 0$.
\end{lemma}

\subsection{Bernstein's Condition}

Further in the paper, we prove the main convergence result for a wide class of distributions satisfying \emph{Bernstein's condition}.

\begin{definition}[{\hspace{-0.01cm}\cite[Eq.~(2.15)]{Wainwright19_1}}] \label{def:bernstein}
	We say that a RV $X \in \Reals$ with mean $\mu$ and variance $\sigma^2$ satisfies \emph{Bernstein's condition} with parameter $\beta > 0$, if
	\[
	\left | \EE{(X- \mu)^k} \right| \leq \frac{1}{2} k! \sigma^2 \beta^{k-2} \quad \text{for $k=2,3,\ldots$}.
	\]
\end{definition}

With some abuse of notation, we write $X \sim \BC{\mu}{\sigma^2}{\beta}$. Note that if $X \sim \BC{\mu}{\sigma^2}{\beta}$, then also $X \sim \BC{\mu}{\sigma^2}{\beta'}$ for any $\beta' \ge \beta$ (monotonicity of the Bernstein parameter).
Examples of  RVs satisfying Bernstein's condition are Gaussian and Laplace RVs, as well as RVs with bounded support. Since for any RV, $|\EE{(X-\mu)^4}| \ge \sigma^4$, it immediately follows that $\nicefrac{\beta}{\sigma} \ge \nicefrac{1}{2\sqrt 3}$.

\begin{lemma} \label{lem:bern-uniform}
The uniform distribution on the interval $[-L + \mu,\mu+L]$ has Bernstein parameter $\beta =  \nicefrac{L}{2\sqrt{5}}$.
\end{lemma}
\begin{IEEEproof}
Omitted for brevity.
\end{IEEEproof}

\begin{lemma}\label{lem:bern-diff2}
	Assume RVs $X_i \sim \BC{\mu_i}{\sigma_i^2}{\beta_i}$, $i=1,\dotsc,n$, are independent, then $X_1 \pm X_2 \pm \dotsb \pm X_n \sim \BC{\mu_1 \pm \mu_2 \pm \dotsb \pm \mu_n}{\sigma^2}{\beta}$, where $\sigma^2 = \sigma_1^2 + \sigma_2^2 + \dotsb + \sigma_n^2$ and $\beta = \min(\beta_1+\beta_2+\dotsb+\beta_n, \sqrt n \max(\sigma_1, \sigma_2, \dotsb, \sigma_n, \beta_1, \beta_2, \dotsb, \beta_n))$.
\end{lemma}
\begin{IEEEproof}
	This lemma is a slight modification of \cite[Prop.~4]{yauhen2024arxiv}.
\end{IEEEproof}

\begin{lemma}[\hspace{-0.01cm}{\cite[Prop. 2.10]{Wainwright19_1}}]
\label{lem:bern-tails}
	If $X \sim \BC{\mu}{\sigma^2}{\beta}$, the following tail bound holds:
	\begin{align*}
		\Prv{|X - \mu| \ge x} &\le 2 \exp \left( -\frac{x^2}{2(\sigma^2 + \beta x)} \right) \\
		&= O\left( \exp \left( - \frac{x}{2 \beta}\right)\right) = o\left( \frac{1}{x^n}\right),
	\end{align*}
	for all $x > 0$ and any positive integer $n$.
\end{lemma}

\subsection{Hypothesis Testing} \label{sec:hypothesis-test-bernstein}
Assume $X \sim \BC{\mu_X}{\sigma_X^2}{\beta_X}$ and $Y \sim \BC{\mu_Y}{\sigma_Y^2}{\beta_Y}$ and define $\sigma^2 = \sigma_X^2 + \sigma_Y^2$ and $\beta = \min(\beta_X+\beta_Y, \sqrt 2 \max(\sigma_X, \sigma_Y, \beta_X, \beta_Y))$. We wish to test:
\begin{align*}
	\mathcal H_0:\; \mu_X = \mu_Y \text{ and } 
	\mathcal H_1:\; \mu_X \neq \mu_Y.
\end{align*}

If $\mu_X = \mu_Y$, but $\mathcal H_0$ is rejected, this is a \emph{type-I error}, and if $\mu_X \neq \mu_Y$, but $\mathcal H_0$ is accepted,  this a \emph{type-II error}. Let the test statistic be $Z = X - Y \sim \BC{\mu_X - \mu_Y}{\sigma^2}{\beta}$. 
Under $\mathcal H_0$ (i.e. when $\mu_X=\mu_Y$), by Lemma~\ref{lem:bern-tails},
\[
\Prv{|Z| \ge z \mid \mathcal H_0} \le 2\exp \left( -\frac{z^2}{2\left(\sigma^2+\beta z\right)} \right).
\]

Now, accept $\mathcal H_0$ if $|Z| < z_\theta$ and reject  $\mathcal H_0$  otherwise, where $\theta$ denotes the desired significance level (i.e., an upper bound on the probability of type-I error). Solving
\[
2\exp \left( -\frac{z_{\theta}^2}{2\left(\sigma^2+\beta z_{\theta}\right)} \right) \le \theta,
\]
we obtain
\[
	z_{\theta} \ge \beta \ln \frac 2\theta + \sqrt{\beta^2 \ln^2 \frac 2\theta + 2\sigma^2 \ln \frac 2\theta}
\]
for $\theta \leq 2\exp(\nicefrac{2\sigma^2}{\beta^2})$. If $\theta \le 2$, and since for any $a,b \ge 0$, $\sqrt{a^2 + b^2} \le a + b$, we can use the simpler value
\[
	z_{\theta} \ge 2\beta \ln \frac 2\theta + \sigma \sqrt{2 \ln \frac 2\theta}.
\]
Both choices guarantee that the probability of type-I error is at most $\theta$.

Under the alternative hypothesis, %
$\mathcal H_1: \mu_X - \mu_Y = \Delta$,
where $\Delta > 0$  w.l.o.g.,  
the test statistic has mean $\EE{Z} = \Delta$ (but the variance and the Bernstein parameter are the same as under $\mathcal H_0$). The probability of type-II error is the probability of accepting $\mathcal H_0$ when $\mathcal H_1$ is true, which can be bounded as follows,
\begin{align*}
	&\Prv{|Z| < z_{\theta} \mid \mathcal H_1} 
	 \le \Prv{|Z - \Delta| > \Delta - z_{\theta} \mid \mathcal H_1} \\	
	&\quad \le 2\exp \left(-\frac{(z_{\theta}-\Delta)^2}{2\left(\sigma^2+\beta (\Delta - z_{\theta})\right)}\right),	
\end{align*}
where the last inequality again follows from Lemma~\ref{lem:bern-tails} applied to the variable $Z - \Delta \sim \BC{0}{\sigma^2}{\beta}$ (under $ \mathcal H_1$).

\section{Model and Algorithm}

\subsection{System Model (Problem Formulation)}
Consider a system of $\numa$ agents connected by a fixed undirected graph $\set G = ([\numa], \set E)$ without self-loops, where $(a,b) \in \set E$ means agents $a$ and $b$ can communicate directly. At synchronized discrete times $t = 1,2,\dotsc$, each agent $a$ privately receives an observation $X_a^{(t)} \in \mathcal{X}_a \subset \mathbb{R}$, where $\mathcal{X}_a $ is bounded. The samples $\{X_a^{(t)}\}_{t \in \mathbb{N}}$ are independent draws from an unknown distribution $\mathcal{D}_a$, whose variance $\sigma_a^2$ is publicly known. As  $\mathcal{X}_a $ is bounded, $\mathcal D_a$ satisfies Bernstein's condition with some parameter $\beta_a$ (publicly known). Each agent aims to estimate the true mean $\mu_a$ of $\mathcal{D}_a$. Although an agent could compute the sample mean of its observations, agents $b \neq a$ may share the same distribution, i.e., $\mathcal{D}_b = \mathcal{D}_a$, so combining data would improve accuracy. However, due to the lack of preliminary information on the agents' distributions and the need for privacy, direct sample exchange is not permitted.

For any agent $a$, define the similarity class $\mathcal C_a \triangleq \{ b \in [\numa] : \mu_b = \mu_a\}$ and let
$\mathcal G_a$ be the connected component of the subgraph of $\mathcal G$ induced by all agents in $\mathcal C_a$, which contains $a$. In other words, it is the subgraph consisting of $a$ and all other agents reachable from $a$ via paths only on the agents from $\mathcal C_a$. The size of $\mathcal G_a$ (number of agents) is denoted by $n_a$.

We propose a collaborative consensus-based algorithm (see \cref{sec:alg}). Each agent $a$ maintains its local sample mean $\bar X_a^{(t)} = \frac{1}{t}\sum_{i=1}^t X_a^{(i)}$ and shares its privatized version, $\tildebar X_a^{(t)}$, with the neighborhood $\set{N}_a = \{ b \in [\numa] : (a,b) \in \set{E} \}$, together with a consensus estimate $\tildehat{\mu}_a^{(t)}$.  %
As $\tildehat{\mu}_a^{(t)}$ is computed from the already protected $\tildebar X_a^{(t)}$, we do not need to explicitly protect $\tildehat{\mu}_a^{(t)}$. %
After each communication round, agents update their consensus estimates using their privatized sample means and (some of) the received consensus estimates. %

Our objective is to design a collaborative algorithm such that, for all sufficiently large $t$, the agents' mean estimates $\hat{\mu}_a^{(t)}$  %
achieve a lower average mean squared error (MSE) than the local estimates (see Proposition~\ref{prop:1}):
\begin{align}
\frac 1M \sum_{a \in [\numa]} \EE{\left(\hat{\mu}_a^{(t)} - \mu_a\right)^2} &< \frac 1M \sum_{a \in [\numa]} \EE{\left(\bar X_a^{(t)} - \mu_a\right)^2} \notag \\
&= \frac 1{Mt} \sum_{a \in [\numa]} \sigma_a^2.
\label{eq:convergence-goal}
\end{align}

As a final remark, while agents could wait until $t \approx t_{\max}$ and then run a consensus algorithm, regular exchanges at each time $t$ are needed to have accurate real-time estimates.

\begin{algorithm}[t]
	\SetAlgoLined
	\KwIn{Graph $\set G = ([\numa], \set E)$ and distributions $\mathcal{D}_a$ for all $a \in [\numa]$} %
	\KwOut{$\hat \mu_a^{(t_{\max})}$ for all $a \in [\numa]$} %
	\BlankLine
	$\tildehat{\mu}_a^{(0)} \gets 0$ for all $a \in [\numa]$ \\ %
	\BlankLine
	\For{$t = 1,2,\dotsc,t_{\max}$}{
		\tcp{In parallel for all  $a \in [\numa]$}
		\tcp{Exchange sample means}
		Obtain $X_a^{(t)} \sim \set D_a$ \\ %
		$\bar X_a^{(t)} \gets \bar X_a^{(t-1)} \times \frac{t-1}{t} + X_a^{(t)} \times \frac{1}{t}$ \\ %
		Sample $Z_a^{(t)} \sim \BC{0}{\sigmaDP^2}{\betaDP}$ \\ %
		$\tilde{\bar{X}}_a^{(t)}\gets \tilde{\bar X}_a^{(t-1)} \times \frac{t-1}{t} + (X_a^{(t)} + Z_a^{(t)}) \times \frac 1t$ \label{line:DPnoise} \\ %
		\ForAll{$b \in \mathcal N_a$ \label{line:sch}}{
			Send $\tilde{\bar X}_a^{(t)}$ \\ %
			Receive $\tilde{\bar X}_b^{(t)}$%
		} 
		\BlankLine
		\tcp{Estimate}
		$\mathcal C_a^{(t)} \gets \{ b \in \mathcal N_a : \chi_a^{(t)}(b; \theta_t)=1 \} \cup \{a\}$\label{line:decision_rule} \\ %
				\tcp{Exchange set sizes estimates}
		\ForAll{$b \in \mathcal N_a$}{
			Send $|\mathcal C_a^{(t)}|$ \label{line:share_C_at}\\ %
			Receive $|\mathcal C_b^{(t)}|$%
		}
		\tcp{Estimate}
		$\tildehat \mu_a^{(t)} \gets (1-\alpha_t) \tildebar X_a^{(t)} + \alpha_t \sum_{b \in \mathcal C_a^{(t)}} W^{(t)}_{ab} \tildehat \mu_b^{(t-1)}$\label{line:mixing} %
		\BlankLine
		\tcp{Exchange consensus means}
		\ForAll{$b \in \mathcal N_a$}{
			Send $\tildehat \mu_a^{(t)}$ \\ %
			Receive $\tildehat \mu_b^{(t)}$ %
		}
		\tcp{If estimated $|\mathcal G_a| \le 2$}
		\MyIf{$|\mathcal C_b^{(t)}| \le 2$ for all $b \in \mathcal C_a^{(t)}$}{
			$\hat \mu_a^{(t)} \gets \bar X_a^{(t)}$ \label{line:na=1} %
		}
		\Else{
			$\hat \mu_a^{(t)} \gets \tildehat \mu_a^{(t)}$ %
		}
	}
	\KwRet{$\hat \mu_a^{(t_{\max})}$ for all $a \in [\numa]$}
	\caption{Private-C-ColME}
	\label{alg:priv-c-colme}
\end{algorithm}

\subsection{Private-C-ColME Algorithm}
\label{sec:alg}
We consider the consensus-based algorithm outlined in \cref{alg:priv-c-colme}, referred to as Private-C-ColME,  which is inspired by \cite[Alg.~1]{galante2024scalable}. The main difference being that in order to provide data privacy, each sample is protected by DP noise in \cref{line:DPnoise} by adding $Z_a^{(t)}  \sim \BC{0}{\sigmaDP^2}{\betaDP}$. Here, $Z_a^{(t)}$ will be taken from the Laplace distribution $\set{L}(0,\nicefrac{\sigmaDP}{\sqrt{2}})$, and for this distribution $\betaDP = \nicefrac{\sigmaDP}{\sqrt{2}}$. Another difference with respect to \cite[Alg.~1]{galante2024scalable} is that we consider an \emph{unrestricted} parallel updating schedule in \cref{line:decision_rule}, i.e., we consider all agents $b \in \mathcal N_a$ and not only the agents in $\set C_a^{(t-1)} \triangleq \{ b \in \mathcal N_a: \chi_a^{(t-1)}(b; \theta_{t-1}) = 1 \} \cup \{a\}$, the estimate of the similarity class $\set C_a$ in the immediate neighborhood by the agent $a$ at time $t-1$, with $\set C_a^{(0)} \triangleq \emptyset$ (i.e., initialized to the empty set).  Hence, if an agent $b$ is removed from agent $a$'s similarity class it can later be added back, which  improves performance. Here, $\chi_a^{(t)}(b; \theta_{t})$ is a decision rule at time $t$ (outlined below), i.e., $\chi_a^{(t)}(b; \theta_t) = 1$ if at time $t$ agent $a$ believes that agent $b$ is in $\set C_a$, and $\theta_t$ is a prescribed  confidence level that  depends on $t$.

Note the special case in the last if-then-else block of \cref{alg:priv-c-colme}. This is an attempt to identify the situations when the estimated size of $\mathcal G_a$ is either $1$ or $2$. In this case, the consensus mean is more noisy then the local estimate (due to DP noise), and the agent reverts to the local estimate.

Further in the paper, we assume two particular choices $\alpha_t = \nicefrac{t}{(t+1)}$ and 
\begin{align} 
W_{ab}^{(t)} = 
\begin{cases}
	\frac{1}{\max \{|\mathcal C_a^{(t)}|, |\mathcal C_b^{(t)}|\} + 1} & \text{if } b \in \mathcal C_a^{(t)} \setminus \{a\}, \\
	1 - \sum_{b \in \mathcal C_a^{(t)} \setminus \{a\}} W_{ab}^{(t)} & \text{if } b = a, \\
	0 & \text{otherwise}
\end{cases}
\label{eq:W}
\end{align}
in \cref{line:mixing}, which are as in  \cite{galante2024scalable}. In order to achieve consensus, we will require
the mixing matrix $W^{(t)}$ to be doubly-stochastic. This is for example satisfied if the decision rule $\chi_a^{(t)}(b; \theta_{t})$ in \cref{line:decision_rule} ensures symmetry: $b \in \mathcal C_a^{(t)}$ if and only if $a \in \mathcal C_b^{(t)}$.  The decision rule is outlined below.

\subsection{Decision Rule}

We consider two decision rules, denoted by $\chi_a^{(t)}(b; \theta_{t})$ and $\tilde{\chi}_a^{(t)}(b; \delta)$, respectively.\footnote{Both decision rules can be used in Algorithm~\ref{alg:priv-c-colme}, although the actual algorithm is typeset using $\chi_a^{(t)}(b; \theta_{t})$ in Line~\ref{line:decision_rule}.}  The first is based on hypothesis testing on Bernstein RVs, while the second is based on \emph{optimistic} distance (see \cite[Eq.~(1)]{galante2024scalable} or \cite[Def.~3]{AsadiBelletMaillardTommasi22_1}).

\subsubsection{Bernstein Rule}
To identify neighbors with the same mean, an agent $a$ runs at each time $t$ individual hypothesis tests against each neighbor $b \in \mathcal N_a$, based on $\tildebar X_a^{(t)}$ and $\tildebar X_b^{(t)}$.\footnote{As $|\mathcal{C}_a^{(t)}|$ is shared in Line~\ref{line:share_C_at}  and also used in the calculation of $\tildehat{\mu}_a^{(t)}$ in Line~\ref{line:mixing} in Algorithm~\ref{alg:priv-c-colme}, which again is shared, also $\bar{X}_a^{(t)}$ must be protected.}  We have (from Lemma~\ref{lem:bern-diff2}) that
\begin{align*}
	\tildebar X_a^{(t)} &\sim \BC{\mu_a}{\frac{\sigma_a^2+\sigmaDP^2}t}{\frac{\tildebar\beta_a}{\sqrt t}}, 
\end{align*}
where
\begin{IEEEeqnarray*}{rCl}
\tildebar\beta_{a}  &\triangleq& \min \left( \max(\sigma_a,\beta_a) + \max(\sigmaDP,\betaDP), \right. \\
&&\hspace*{1.25cm} \> \left. \max\left(\beta_a+ \betaDP,\sqrt{\sigma^2_a+\sigmaDP^2}\right)\right),
\end{IEEEeqnarray*}
and similarly for agent $b$.
We apply the hypothesis test from \cref{sec:hypothesis-test-bernstein}. More precisely, for a desired confidence level $\theta_t$, we choose 
\begin{align*}
	&z_{\theta_t} %
	 = 2 \frac{\tildebar\beta_a + \tildebar\beta_b}{\sqrt t} \ln \frac 2{\theta_t} + \frac{\sqrt{\sigma_a^2 + \sigma_b^2 + 2\sigmaDP^2}}{\sqrt t} \sqrt{2 \ln \frac 2{\theta_t}} 
\end{align*}
and we define, for $b \neq a$,
\[
	\chi_a^{(t)}(b ; \theta_t) =
	\begin{cases}
		1 & \text{if } \left| \tildebar X_a^{(t)} - \tildebar X_b^{(t)} \right| < z_{\theta_t},\\ %
		0 & \text{otherwise.}
	\end{cases}
\]
For large $t$, $\tildebar X_a^{(t)} - \tildebar X_b^{(t)}$ concentrates around $\mu_a - \mu_b$, which is $0$ for $\mu_a = \mu_b$, and $\Delta = \mu_a - \mu_b > 0$ otherwise. Therefore, we want $z_{\theta_t} \to 0$ with $t \to \infty$, which separates $0$ and $\Delta > 0$. This is satisfied when 
	$\frac{1}{\sqrt t} \ln \frac 1{\theta_t} \to 0$,
which is equivalent to $\nicefrac{1}{\theta_t} = \mathrm e^{o(\sqrt t)}$. On the other hand, we want $\theta_t \to 0$ so that the probability of type-I error vanishes. Intuitively, we want $\theta_t$ to decay to $0$ but not too fast.

For large $t$, the nominator in the type-II error probability is $(o(1) - \Delta)^2 = \Delta^2 + o(1)$, and the denominator is
\begin{align*}
&2\left(\frac{\sigma_a^2 + \sigma_b^2 + 2\sigmaDP^2}t + \frac{\tildebar\beta_a + \tildebar\beta_b}{\sqrt t}(\Delta - o(1))\right) \\
&\quad= \frac{2(\tildebar\beta_a+\tildebar\beta_b)\Delta}{\sqrt t} + o\left( \frac{1}{\sqrt t}\right).
\end{align*}

Therefore, the type-II error probability is asymptotically not more than%
\[
	\exp \left( -\frac{\Delta \sqrt t}{2(\tildebar\beta_a + \tildebar\beta_b)} \right).
\]

\subsubsection{Optimistic Distance}
The decision rule in \cite[Alg.~1]{galante2024scalable} is based on \emph{optimistic} distance (see \cite[Eq.~(1)]{galante2024scalable}, also considered in \cite[Def.~3]{AsadiBelletMaillardTommasi22_1}) where the confidence bound (the $\beta_\delta$-parameter) is set according to  \cite[Eq.~(3)]{galante2024scalable} (or \cite[Lem.~1]{AsadiBelletMaillardTommasi22_1}).
Here, as in  \cite{galante2024scalable}, the $\gamma$ in \cite[Eq.~(3)]{galante2024scalable} is set equal to $\nicefrac{\delta}{4rM}$, where $r$ is the assumed regularity of the graph $\mathcal{G}$ and $\delta \in (0,1]$. 
We adjust the rule for DP noise and denote it $\tilde{\chi}_a^{(t)}(b; \delta)$ in the following.\footnote{Note that \cite[Lem.~1]{AsadiBelletMaillardTommasi22_1} assumes that $X_a^{(t)} + Z_a^{(t)}$ follows a sub-Gaussian distribution. Here, we use this rule as a ``heuristic'', as sub-Gaussianity does not hold when  $Z_a^{(t)}$ follows a Laplace distribution. In  \cite[Eq.~(5)]{galante2024scalable}, an expression for the confidence bound valid for bounded fourth-central-moment distributions for $X_a^{(t)} + Z_a^{(t)}$ is given. However, this expression gives a worse  performance in our setup compared \cite[Eq.~(3)]{galante2024scalable} (which assumes  sub-Gaussianity) due to a looser bounding argument in its proof.\looseness-1} In particular, 
we let $\tilde{\chi}_a^{(t)}(b; \delta)=1$, for $b \neq a$, if
\begin{align*}
\left| \tildebar{X}_a^{(t)} - \tilde{\bar{X}}_b^{(t)}   \right| - \tilde{\beta}_{\delta}(a;t) -  \tilde{\beta}_{\delta}(b;t) \leq 0
\end{align*}
and $0$, otherwise, where $\delta \in (0,1]$ and 
\begin{align*}
&\tilde{\beta}_{\delta}(\cdot; t) 
= 
\sqrt{\frac{2(\sigmaDP^2 + \sigma_{\cdot}^2)}{t} \left( 1 + \frac{1}{t} \right) \ln \left( \frac{4r\numa \sqrt{t+1}}{ \delta}  \right)   }. %
\end{align*}%

 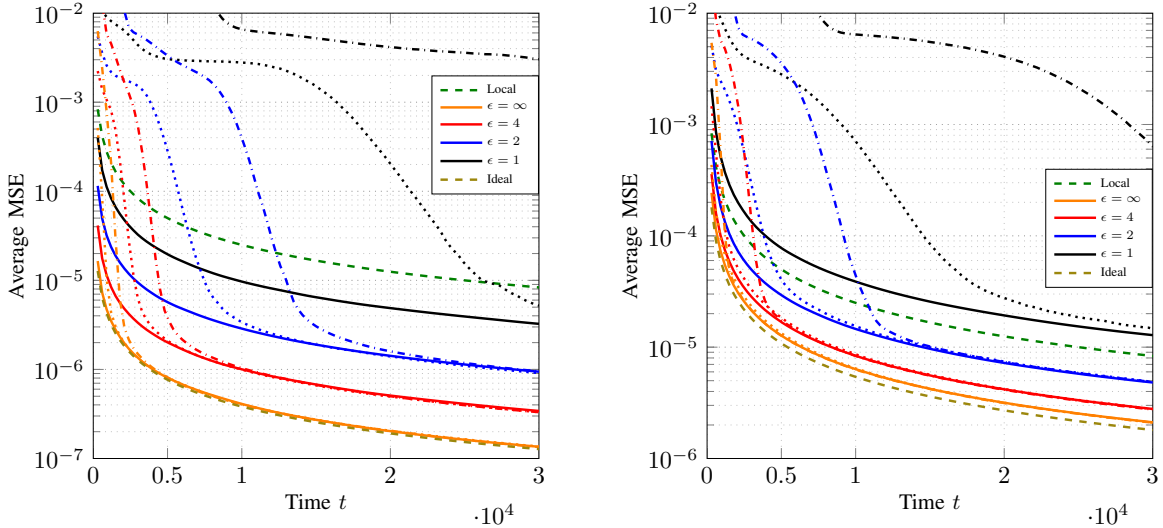
\begin{figure*}[t!]
		\centering
		\subfloat{
			\definecolor{darkyellow}{rgb}{0.61,0.53,0.05}

\begin{tikzpicture}[thick, scale=0.925, every node/.style={transform shape}]
\begin{axis}[%
width=0.9\columnwidth,
height=0.9\columnwidth,
xmin=1.0,
xmax=30000,
xtick={1,10000,20000,30000,40000,5000,60000,70000,80000,90000,100000},
xlabel={Time $t$},
xticklabel style = {/pgf/number format/fixed, /pgf/number format/precision=2},
xlabel style={
	yshift=0.5ex,
	name=label,
	font=\small},
grid style={gray,opacity=0.5,dotted},
xmajorgrids,
ymajorgrids,
yminorgrids,
ymode=log,
ymin=1e-7,
ymax=1e-2,
ytick={1e-7,1e-6,1e-5,1e-4,1e-3,1e-2,1e-1},
ylabel={Average MSE},
legend style={font=\scriptsize},
ylabel style={
	yshift=-1.0ex,
	name=label,
	font=\small},
axis background/.style={fill=white},
legend cell align=left,
legend style={at={(axis cs: 30000,2e-3)},anchor=north east,nodes={scale=0.75, transform shape}},
]

\addplot [color=darkgreen,dashed,line width=1pt,mark options={solid, line width = 0.5pt, fill=white}]table[x=time,y=acc_local] {data_final_version/local_curve.txt.100};
\addlegendentry{Local};

\addplot [color=orange, solid,line width=1pt, mark options={solid, line width = 0.5pt, fill=white}]table[x=time,y=MSE]
{data_final_version/ImpAlg2LaplaceOracleParallel_PM1_UniformData_RR_curve_M200_NumClasses_3_eps1000000.0000_delta0.000000_r20_scaletime_10_.txt.100};
\addlegendentry{$\epsilon=\infty$};

\addplot [color=red, solid,line width=1pt, mark options={solid, line width = 0.5pt, fill=white}]table[x=time,y=MSE]
{data_final_version/ImpAlg2LaplaceOracleParallel_PM1_UniformData_RR_curve_M200_NumClasses_3_eps4.0000_delta0.000000_r20_scaletime_10_.txt.100};
\addlegendentry{$\epsilon=4$};

\addplot [color=blue, solid,line width=1pt, mark options={solid, line width = 0.5pt, fill=white}]table[x=time,y=MSE]
{data_final_version/ImpAlg2LaplaceOracleParallel_PM1_UniformData_RR_curve_M200_NumClasses_3_eps2.0000_delta0.000000_r20_scaletime_10_.txt.100};
\addlegendentry{$\epsilon=2$};

\addplot [color=black, solid,line width=1pt, mark options={solid, line width = 0.5pt, fill=white}]table[x=time,y=MSE]
{data_final_version/ImpAlg2LaplaceOracleParallel_PM1_UniformData_RR_curve_M200_NumClasses_3_eps1.0000_delta0.000000_r20_scaletime_10_.txt.100};
\addlegendentry{$\epsilon=1$};

\addplot [color=orange, dotted,line width=1pt, forget plot, mark options={solid, line width = 0.5pt, fill=white}]table[x=time,y=MSE]
{data_final_version/ImpAlg2LaplaceNewBernsteinTwoParallel_PM1_UniformData_RR_curve_M200_NumClasses_3_eps1000000.0000_delta0.000000_r20_scaletime_10_.txt.100};

\addplot [color=red, dotted,line width=1pt, forget plot,mark options={solid, line width = 0.5pt, fill=white}]table[x=time,y=MSE]
{data_final_version/ImpAlg2LaplaceNewBernsteinTwoParallel_PM1_UniformData_RR_curve_M200_NumClasses_3_eps4.0000_delta0.000000_r20_scaletime_10_.txt.100};

\addplot [color=blue, dotted,line width=1pt, forget plot,mark options={solid, line width = 0.5pt, fill=white}]table[x=time,y=MSE]
{data_final_version/ImpAlg2LaplaceNewBernsteinTwoParallel_PM1_UniformData_RR_curve_M200_NumClasses_3_eps2.0000_delta0.000000_r20_scaletime_10_.txt.100};
\addplot [color=black, dotted,line width=1pt, forget plot, mark options={solid, line width = 0.5pt, fill=white}]table[x=time,y=MSE]
{data_final_version_finetune/ImpAlg2LaplaceNewBernsteinTwoParallel_PM1_UniformData_RR_curve_M200_NumClasses_3_eps1.0000_delta0.000000_r20_scaletime_10_.txt.100};

\addplot [color=orange, dashdotted,line width=1pt,  forget plot,mark options={solid, line width = 0.5pt, fill=white}]table[x=time,y=MSE]
{data_final_version_finetune/ImpAlg2LaplaceNewOptDist_1.0000Parallel_PM1_UniformData_RR_curve_M200_NumClasses_3_eps1000000.0000_delta0.000000_r20_scaletime_10_.txt.100};

\addplot [color=red, dashdotted,line width=1pt, forget plot,mark options={solid, line width = 0.5pt, fill=white}]table[x=time,y=MSE]
{data_final_version_finetune/ImpAlg2LaplaceNewOptDist_1.0000Parallel_PM1_UniformData_RR_curve_M200_NumClasses_3_eps4.0000_delta0.000000_r20_scaletime_10_.txt.100};

\addplot [color=blue, dashdotted,line width=1pt, forget plot,mark options={solid, line width = 0.5pt, fill=white}]table[x=time,y=MSE]
{data_final_version_finetune/ImpAlg2LaplaceNewOptDist_1.0000Parallel_PM1_UniformData_RR_curve_M200_NumClasses_3_eps2.0000_delta0.000000_r20_scaletime_10_.txt.100};

\addplot [color=black, dashdotted,line width=1pt,  forget plot,mark options={solid, line width = 0.5pt, fill=white}]table[x=time,y=MSE]
{data_final_version_finetune/ImpAlg2LaplaceNewOptDist_1.0000Parallel_PM1_UniformData_RR_curve_M200_NumClasses_3_eps1.0000_delta0.000000_r20_scaletime_10_.txt.100};

\addplot [color=darkyellow,dashed,line width=1pt,mark options={solid, line width = 0.5pt, fill=white}]table[x=time,y=error] {data_final_version/ideal_curve_consensus_r20_M200.txt.100};
\addlegendentry{Ideal};

\end{axis}
\end{tikzpicture}%
		}
		\hspace{2ex}
		\subfloat{
			\definecolor{darkyellow}{rgb}{0.61,0.53,0.05}

\begin{tikzpicture}[thick, scale=0.925, every node/.style={transform shape}]
\begin{axis}[%
width=0.9\columnwidth,
height=0.9\columnwidth,
xmin=1.0,
xmax=30000,
xtick={1,10000,20000,30000,40000,5000,60000,70000,80000,90000,100000},
xlabel={Time $t$},
xticklabel style = {/pgf/number format/fixed, /pgf/number format/precision=2},
xlabel style={
	yshift=0.5ex,
	name=label,
	font=\small},
grid style={gray,opacity=0.5,dotted},
xmajorgrids,
ymajorgrids,
yminorgrids,
ymode=log,
ymin=1e-6,
ymax=1e-2,
ytick={1e-7,1e-6,1e-5,1e-4,1e-3,1e-2,1e-1},
ylabel={Average MSE},
legend style={font=\scriptsize},
ylabel style={
	yshift=-1.0ex,
	name=label,
	font=\small},
axis background/.style={fill=white},
legend cell align=left,
legend style={at={(axis cs: 30000,4e-4)},anchor=north east,nodes={scale=0.75, transform shape}},
]

\addplot [color=darkgreen,dashed,line width=1pt,mark options={solid, line width = 0.5pt, fill=white}]table[x=time,y=acc_local] {data_final_version/local_curve.txt.100};
\addlegendentry{Local};

\addplot [color=orange, solid,line width=1pt, mark options={solid, line width = 0.5pt, fill=white}]table[x=time,y=MSE]
{data_final_version/ImpAlg2LaplaceOracleParallel_PM1_UniformData_RR_curve_M200_NumClasses_3_eps1000000.0000_delta0.000000_r5_scaletime_10_.txt.100};
\addlegendentry{$\epsilon=\infty$};

\addplot [color=red, solid,line width=1pt, mark options={solid, line width = 0.5pt, fill=white}]table[x=time,y=MSE]
{data_final_version/ImpAlg2LaplaceOracleParallel_PM1_UniformData_RR_curve_M200_NumClasses_3_eps4.0000_delta0.000000_r5_scaletime_10_.txt.100};
\addlegendentry{$\epsilon=4$};

\addplot [color=blue, solid,line width=1pt, mark options={solid, line width = 0.5pt, fill=white}]table[x=time,y=MSE]
{data_final_version/ImpAlg2LaplaceOracleParallel_PM1_UniformData_RR_curve_M200_NumClasses_3_eps2.0000_delta0.000000_r5_scaletime_10_.txt.100};
\addlegendentry{$\epsilon=2$};

\addplot [color=black, solid,line width=1pt, mark options={solid, line width = 0.5pt, fill=white}]table[x=time,y=MSE]
{data_final_version/ImpAlg2LaplaceOracleParallel_PM1_UniformData_RR_curve_M200_NumClasses_3_eps1.0000_delta0.000000_r5_scaletime_10_.txt.100};
\addlegendentry{$\epsilon=1$};

\addplot [color=orange, dotted,line width=1pt, forget plot, mark options={solid, line width = 0.5pt, fill=white}]table[x=time,y=MSE]
{data_final_version/ImpAlg2LaplaceNewBernsteinTwoParallel_PM1_UniformData_RR_curve_M200_NumClasses_3_eps1000000.0000_delta0.000000_r5_scaletime_10_.txt.100};

\addplot [color=red, dotted,line width=1pt, forget plot,mark options={solid, line width = 0.5pt, fill=white}]table[x=time,y=MSE]
{data_final_version/ImpAlg2LaplaceNewBernsteinTwoParallel_PM1_UniformData_RR_curve_M200_NumClasses_3_eps4.0000_delta0.000000_r5_scaletime_10_.txt.100};

\addplot [color=blue, dotted,line width=1pt, forget plot,mark options={solid, line width = 0.5pt, fill=white}]table[x=time,y=MSE]
{data_final_version/ImpAlg2LaplaceNewBernsteinTwoParallel_PM1_UniformData_RR_curve_M200_NumClasses_3_eps2.0000_delta0.000000_r5_scaletime_10_.txt.100};
\addplot [color=black, dotted,line width=1pt, forget plot, mark options={solid, line width = 0.5pt, fill=white}]table[x=time,y=MSE]
{data_final_version_finetune/ImpAlg2LaplaceNewBernsteinTwoParallel_PM1_UniformData_RR_curve_M200_NumClasses_3_eps1.0000_delta0.000000_r5_scaletime_10_.txt.100};

\addplot [color=orange, dashdotted,line width=1pt,  forget plot,mark options={solid, line width = 0.5pt, fill=white}]table[x=time,y=MSE]
{data_final_version_finetune/ImpAlg2LaplaceNewOptDist_1.0000Parallel_PM1_UniformData_RR_curve_M200_NumClasses_3_eps1000000.0000_delta0.000000_r5_scaletime_10_.txt.100};

\addplot [color=red, dashdotted,line width=1pt, forget plot,mark options={solid, line width = 0.5pt, fill=white}]table[x=time,y=MSE]
{data_final_version_finetune/ImpAlg2LaplaceNewOptDist_1.0000Parallel_PM1_UniformData_RR_curve_M200_NumClasses_3_eps4.0000_delta0.000000_r5_scaletime_10_.txt.100};

\addplot [color=blue, dashdotted,line width=1pt, forget plot,mark options={solid, line width = 0.5pt, fill=white}]table[x=time,y=MSE]
{data_final_version_finetune/ImpAlg2LaplaceNewOptDist_1.0000Parallel_PM1_UniformData_RR_curve_M200_NumClasses_3_eps2.0000_delta0.000000_r5_scaletime_10_.txt.100};

\addplot [color=black, dashdotted,line width=1pt,  forget plot,mark options={solid, line width = 0.5pt, fill=white}]table[x=time,y=MSE]
{data_final_version_finetune/ImpAlg2LaplaceNewOptDist_1.0000Parallel_PM1_UniformData_RR_curve_M200_NumClasses_3_eps1.0000_delta0.000000_r5_scaletime_10_.txt.100};

\addplot [color=darkyellow,dashed,line width=1pt,mark options={solid, line width = 0.5pt, fill=white}]table[x=time,y=error] {data_final_version/ideal_curve_consensus_r5_M200.txt.100};
\addlegendentry{Ideal};

\end{axis}
\end{tikzpicture}%
		}
		\vspace{-2ex}
		\caption{Comparing the average MSE of Private-C-ColME  for different privacy levels $\epsilon$ and three different decision rules: oracle (solid curves), Bernstein hypothesis testing (dotted), and optimistic distance (dashdotted). There are $M=200$  agents forming  three classes with $r=20$ (left-most plot) and $r=5$ (right-most plot). 
			The curves are for uniform data with  standard deviation  $\sigma=\nicefrac{1}{2}$ and $L = \sigma \sqrt{3}$. The results are based on $4000$ simulation runs.}%
		\label{fig:1}
		\vspace{-2ex}
\end{figure*}

\section{Privacy and Convergence}
Here, we show that \cref{alg:priv-c-colme} guarantees DP and also converges faster than a fully local approach (cf. \cref{eq:convergence-goal}), under an oracle decision rule and under some conditions on the privacy level $\epsilon$ and the  graph $\set G$.

\subsection{Privacy}
Each element $X_a^{(t)} \in \mathcal{X}_a$ is protected by an individual noise term $Z_a^{(t)} \sim \BC{0}{\sigmaDP^2}{\betaDP}$, and  all other calculations that are exchanged with the neighbors use this privatized element, $X_a^{(t)} + Z_a^{(t)}$. Here, $Z_a^{(t)}$ will be taken from the Laplace distribution $\set{L}(0,\nicefrac{\sigmaDP}{\sqrt{2}})$, and hence, according to Lemma~\ref{lem:DP_Laplace} (with $n=1$), \cref{alg:priv-c-colme} provides agent-level  data DP.\looseness-1%
 
\subsection{Convergence}
We want to show that we converge faster than the fully local approach, as it is stated in \cref{eq:convergence-goal}. We want to prove the convergence of the public consensus estimate \emph{when an oracle is used for the decision rule}, i.e., when for all $t$ and all $a$, $\mathcal C_a^{(t)} = \{a\} \cup (\mathcal N_a \cap \mathcal C_a)$. The intuition is that if the decision rule is asymptotically correct (with vanishing type-I and type-II error probabilities), at some point \cref{alg:priv-c-colme} reaches the state when all the agents only take into account the messages from their neighbors from their own class, and the theorem applies.\looseness-1

\begin{theorem}\label{thm:convergence}
	For any distributions $\mathcal D_a$, $a \in [M]$, with bounded support, the average MSE of \cref{alg:priv-c-colme} with $\alpha_t = \nicefrac{t}{(t+1)}$, mixing matrix $W^{(t)}$ from \cref{eq:W}, and oracle decision rule is
	\begin{align*}
		&\frac 1{\numa} \sum_{a \in [\numa]} \EE{\left( \hat \mu_a^{(t)} - \mu_a \right)^2} 
		\\
		& = \frac 1{\numa t} \Bigl( \sum_{\substack{a \in [\numa]:\\n_a \le 2}} \sigma_a^2 +  \sum_{\substack{a \in [\numa]:\\n_a \ge 3}} \frac{2(\sigma_a^2 + \sigmaDP^2)}{n_a} \Bigr) + o\left( \frac 1t \right).%
	\end{align*}
\end{theorem}

\begin{IEEEproof}[Proof Sketch]
The mixing matrix $W^{(t)}$ in \cref{eq:W} is independent of $t$ under the oracle decision rule, hence we denote it by $W$. %
	The proof unrolls the recursion of the error and shows it contracts under $W$, whose powers converge to a rank-one projection due to the spectral gap.  The dominant term in the error of order $O(\nicefrac 1t)$ comes from accumulated noise, while the lower-order  term is bounded using decay of $W^t$. %
\end{IEEEproof}

\begin{cor}\label{cor:sigmaDP-bound}
	In the settings of Theorem~\ref{thm:convergence}, let the DP noise variance satisfy
	\begin{align}
		\sigmaDP^2 < \frac{\sum_{a \in [M]:n_a \ge 3}\sigma_a^2 \left( 1 - \nicefrac 2{n_a}\right)}{2\sum_{a \in [M]:n_a \ge 3} \nicefrac{1}{n_a}}. \label{eq:thm1}
	\end{align}
	Then, \cref{alg:priv-c-colme} converges faster than the local approach.
\end{cor}

\subsection{Benchmarks}

\begin{proposition}[Local] \label{prop:1}
The average MSE of a pure local approach is
\begin{align*} %
\frac{1}{\numa} \sum_{a \in [\numa]}\EE{\left(\hat \mu_a^{(t)} - \mu_a\right)^2} =  \frac{1}{\numa t}\sum_{a \in [\numa]}\sigma_a^2.
\end{align*}
\end{proposition}

On the other hand, if privacy is ignored, and agent $a$ knows $\mathcal G_a$ and has access to \emph{all} the data of all the agents in $\mathcal G_a$, %
it virtually has one large sample of size $n_a t$ (as opposed to the local sample of size $t$), which is ideal.  %
\begin{proposition}[Ideal] \label{prop:3}
The average MSE of an \emph{ideal} scheme is
\begin{align*}
\frac{1}{\numa} \sum_{a \in [\numa]}\EE{\left(\hat \mu_a^{(t)} - \mu_a\right)^2} = \frac{1}{\numa t}\sum_{a \in [\numa]}\frac{\sigma_a^2}{n_a},
\end{align*}
and no approach can perform better than this.
\end{proposition}

\section{Numerical Experiments}

We consider the case of $\numa$ agents from three classes. The  agents are placed uniformly at random within the  classes, giving roughly balanced class sizes.  The agents'  data distributions are uniform (to model tabular data) on a range of size $2L=2 \sigma  \sqrt{3}$ (giving a standard deviation of $\sigma$), with $\sigma=\nicefrac{1}{2}$, but different class-dependent means;  $\nicefrac{1}{5}$, $\nicefrac{2}{5}$, and $\nicefrac{4}{5}$. %
The underlying communication graph is a random $r$-regular graph (without self-loops) and we consider Laplace DP noise. Simulations of \cref{alg:priv-c-colme} under Theorem~\ref{thm:convergence} matched the theory; corresponding theoretical curves are omitted due to space. Further, we  use %
$\alpha_t= \nicefrac{\lfloor t/10 \rfloor+1}{\lfloor t /10 \rfloor+2}$, which gives  better performance (even asymptotically) than the prediction of Theorem~\ref{alg:priv-c-colme}. As in \cite{galante2024scalable} (see Appendix~I.3),  $\alpha_t$ is refreshed at each topology change: when $\mathcal{C}_a^{(t)} \neq \mathcal{C}_a^{(t-1)}$, the time $t$ is reset to $1$ when calculating $\alpha_t$. This is done individually for each agent $a$ and hence effectively gives different $\alpha$'s for different agents $a$.\looseness-1

In \cref{fig:1}, we compare the performance for different privacy levels $\epsilon$ (including no privacy, i.e.,  $\epsilon=\infty$) for $\numa=200$ and $r=20$ (left-most plot) and $r=5$ (right-most plot) under two  decision rules; optimistic distance (as studied in \cite{galante2024scalable}; dashdotted curves) and  %
hypothesis testing based on Bernstein RVs (dotted curves). 
For the hypothesis testing decision rule, for $r=5$, we use for $\theta_t$ the minimum of $2$ and  $\nicefrac{3}{t^{\nicefrac{1}{8}}}$ for $\epsilon=1$ and $2$ and $\nicefrac{3}{t^{\nicefrac{1}{7}}}$ for $\epsilon=2$, $4$, and $\infty$, while  for $r=20$, we use for $\theta_t$ the minimum of $2$ and $\nicefrac{3}{t^{\nicefrac{1}{7}}}$ for $\epsilon=1$,  $2$ and $\nicefrac{3}{t^{\nicefrac{1}{6}}}$ for $\epsilon=2$, and  $2$ and $\nicefrac{3}{t^{\nicefrac{1}{5}}}$ for $\epsilon=4$ and $\infty$. 
The values for $\theta_t$ are  fine-tuned and for $\beta_a$ for all agents $a$ we use Lemma~\ref{lem:bern-uniform}. Moreover, as the DP noise is distributed according to the Laplace distribution, $\betaDP = \nicefrac{\sigmaDP}{\sqrt{2}}$. For optimistic distance we use $\delta=1$.\footnote{In \cite{galante2024scalable},   $\delta=\nicefrac{1}{10}$ was used, but we have observed better performance in our setup with $\delta=1$ for all values of $\epsilon$.}
For comparison, we also show the performance with an oracle decision rule (solid curves), i.e., all agents know at any time which neighboring agents are in their class. As can be seen from the figure, collaboration may give a benefit, i.e., the average MSE can be lower than that of a pure local approach (see Proposition~\ref{prop:1}; dashed green curve) asymptotically, i.e., when the error is sufficiently low, depending on the privacy level $\epsilon$ and the connectivity $r$. 
For $r=5$, the right-hand side of the bound in \eqref{eq:thm1} in Corollary~\ref{cor:sigmaDP-bound} is approximately $1.9$ when averaged over the random assignment of agents to classes and communication graphs $\set{G}$, while for $r=20$ this number is approximately $8.1$.
This  explains why there is no collaborative gain for $\epsilon=1$ for $r=5$ as the corresponding $\sigmaDP^2$ does not satisfy the bound in Corollary~\ref{cor:sigmaDP-bound}, while for all other cases the bound is satisfied.  %
As is apparent from Corollary~\ref{cor:sigmaDP-bound}, a lower connectivity (and hence a lower $n_a$) requires a lower privacy level (i.e., higher $\epsilon$ and lower $\sigmaDP$) in order to have a collaborative gain. 
Second, the hypothesis testing decision rule outperforms  optimistic distance for both $r=5$ and $r=20$, and for all privacy levels.  %
Third,  the performance with both decision rules approaches the oracle decision rule benchmark as the time $t$ increases, except possibly for optimistic distance for $\epsilon=1$, which is according to  intuition that the decision rules are asymptotically correct. 
 Fourth, there is  some performance gap to ideal performance (dashed dark yellow curve), which is the performance when all agents $a$ have access to \emph{all} the data of all the connected agents in $\mathcal C_a$ at any time $t$ (see Proposition~\ref{prop:3}), when we impose a privacy constraint.  %
However, with no privacy, \cref{alg:priv-c-colme} performs very close to ideal performance for low error rates.

\IEEEtriggeratref{11}

%
%
%
%
%
%
%
%
\end{document}